\newcommand{\q}{\vec q}
\newcommand{\x}{\vec x}
\newcommand{\xp}{\vec x^{\, \prime}}
\newcommand{\xpp}{\vec x^{\,\prime\prime }}
\newcommand{\xppp}{\vec x^{\, \prime\prime\prime }}

\newcommand{\im}{\rm  i}

\documentclass[prl,aps,epsfig]{revtex4}

\usepackage{graphics}
\begin{document}
\title{Correlated imaging, quantum and classical}
\author{A.~Gatti, E.~Brambilla, M. Bache and L.~A.~Lugiato}
\address{INFM, Dipartimento di Scienze CC.FF.MM.,
Universit\`a dell'Insubria, Via Valleggio 11, 22100 Como, Italy}
\begin{abstract}
We analytically show that it is possible to perform coherent imaging by using the classical correlation 
of two beams obtained by splitting incoherent thermal radiation. The case 
of such two classically correlated beams is treated in parallel with 
the configuration based on two entangled beams  produced
by parametric down-conversion, and a basic analogy is pointed out. The results are compared in a specific numerical example.
\end{abstract}
\centerline{Version \today}
\maketitle
The topic of entangled imaging  has attracted noteworthy attention in recent years\cite{bib1,bib2,bib3,bib4,bib5,bib7,bib13,bib14}.
This tecnique exploits
the quantum entaglement of the state generated by parametric down-conversion (PDC), in order to retrieve information about
an unknown object. 
In the regime of single photon pair production of PDC, the  photons of a pair 
are spatially separated and propagate through two distinct imaging systems. In the path of one of the photons
an object is located. Information about the spatial distribution of the object is not obtained by detection of this photon,
but rather by registering
the coincidence counts as a function of the other photon position \cite{bib1,bib2,bib3,bib4,bib5}.  
In the regime of a large number of photon pairs,
this procedure is generalized to the measurement of the signal-idler
spatial correlation function of intensity fluctuations \cite{bib7}. Such a two-arm configuration provides more
flexibility in comparison with standard imaging procedures,
as e.g. the possibility of illuminating the object
with one light frequency
and performing a spatially resolved detection in the other arm with a different light frequency,
or of processing the information from the object by only operating on the imaging system of arm 2 \cite{bib5,bib7}. 
In addition, it opens the possibility of performing coherent imaging by using, in a sense,  spatially incoherent light,
since each of the two down-converted beams taken separately is described by a thermal-like mixture
and only the two-beam state
is pure(see e.g. \cite{bib5} and \cite{bib7}).\\
In this paper we show that it is possible to implement such a scheme using a truly
incoherent light, as the radiation produced by a thermal (or quasi-thermal) source.  
A comparison  between thermal and biphoton emission is performed in
\cite{bib8b}, where an underlying duality accompanies the mathematical similarity between the two cases.
\begin{figure}[h]
{\scalebox{.4}{\includegraphics*{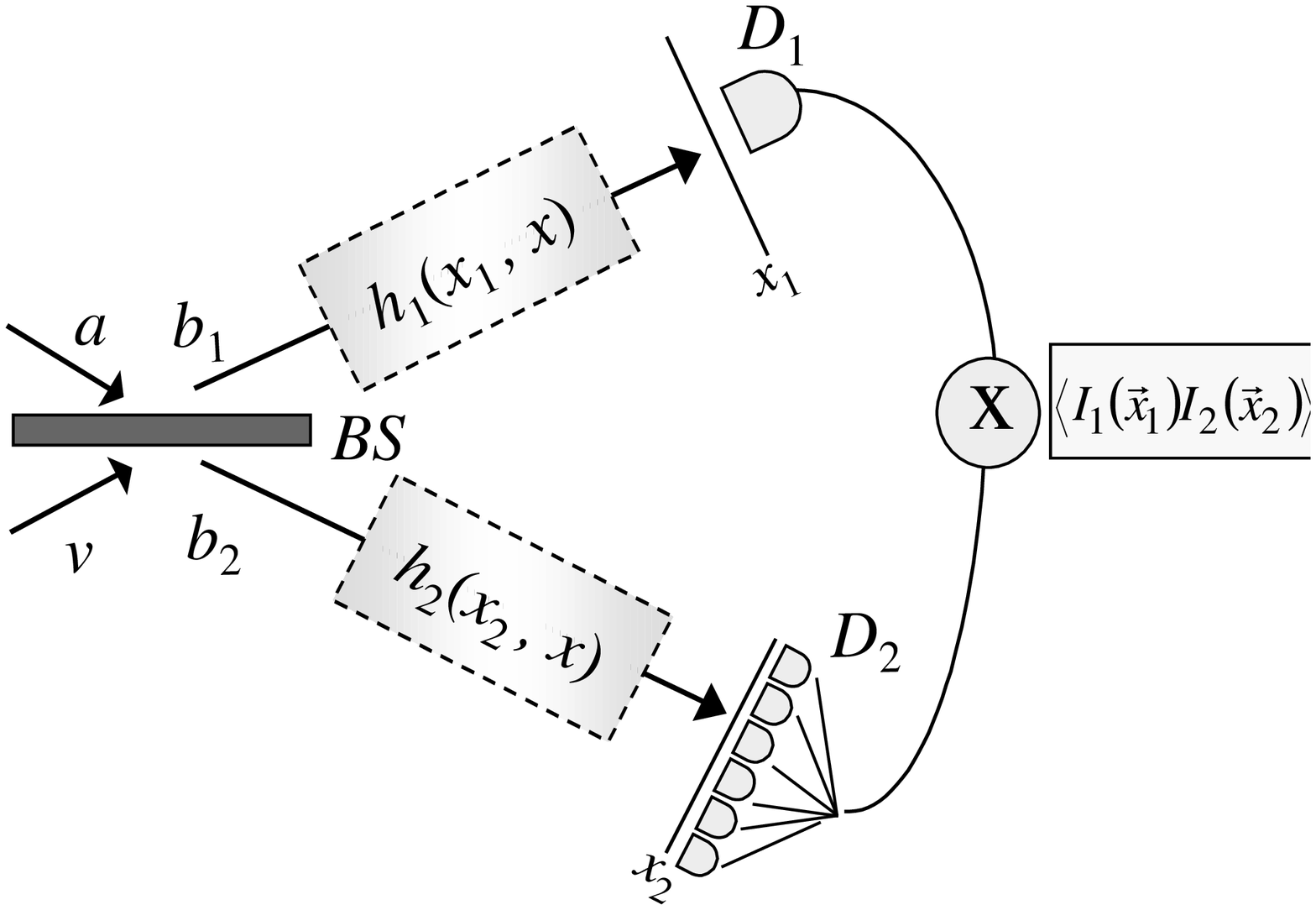}}}
\caption{Correlated imaging with incoherent thermal light. The thermal beam $a$ is splitted
into two beams which travel through two distinct imaging systems,
described by their impulse response functions $h_1$ and $h_2$. Arm 1 includes an object.
Detector $D_1$ is either a point-like detector or a bucket detector. Beam 2 is detected by
an array of pixel detectors.
$v$ is a vacuum field.}
\label{fig2}
\end{figure}
 Here, we consider a different scheme (Fig.\ref{fig2}), appropriate for correlated imaging,
in which a thermal 
beam is divided  by a beam-splitter (BS) and the two outgoing beams are handled in the same way as the
PDC beams in entangled imaging. A basic analogy between the PDC and the thermal case emerges from our analysis.\\
Currently there is a debate whether quantum entanglement is necessary to perform correlated imaging
\cite{bib5,bib13,bib7,bib14}. The discussion became very lively after the experiment of \cite{bib13}
reproduced the results of a ghost image experiment \cite{bib3} by using classically correlated beams.
We will show here that the spatial correlation 
of the two beams produced by splitting thermal light, although being completely classical, is enough to qualitatively
 reproduce 
all the features of the entangled imaging.

For the sake of comparison we will treat in parallel the cases of entangled
beams and of thermal light. For simplicity,
we consider only spatial variables and ignore the time argument, which corresponds to using
a narrow frequency filter. We will come back to this point in the final part of the paper.
In addition, we assume translational invariance in the transverse plane, which 
amounts to requiring that the cross-section
of the source is much larger than the object and all the optical elements.
In a future publication 
we will release these assumptions.

In the entangled  case, the signal and idler fields are generated 
in a type II $\chi^{(2)}$ crystal by a PDC process. Our starting point  are the
input-output relations of the crystal, which in the plane-wave pump approximation read \cite{bib7,bib8,bib9}
\begin{equation}
b_i (\q)= U_i(\q) a_i (\q) + V_i (\q) a_j^\dagger (-\q) \quad i \ne j =1,2 \, .
                    \label{eq1}
\end{equation}
Here,
$b_i (\q) = \int \frac{{\rm d} \x}{2 \pi}  e^{-\im \q\cdot \x} b_i (\x) $, 
where $b_i(\x)$, are the signal $(i=1)$  and idler $(i=2)$
field envelope operators at the output face of the crystal (distinguished 
by their orthogonal polarizations), $\x$ being
the position in the transverse plane. $
a_i \, , \; i=1,2 $ are the corresponding fields at the input face of the crystal, and are in the vacuum state.
The gain functions $U_i, V_i$ are for example given in \cite{bib8}. 

In the thermal case, we start from the input/output relations of a beam splitter
\begin{eqnarray}
b_1 (\x )= r a(\x) + t v (\x) \, ,\quad
b_2 (\x )= t a (\x)+ r v(\x) 
                                        \label{BS} \: ,
\end{eqnarray}
where $t$ and $r$ are the transmission and reflection coefficients of the mirror, $a$ is a thermal field 
and
$v$ is a vacuum field  uncorrelated from $a$. We assume that the thermal state of $a$ is
characterized by 
a Gaussian field statistics, in which  any correlation function of arbitrary order
is expressed via the second order correlation function \cite{bib10}:
\begin{equation}
\langle a^\dagger (\x) a (\xp) \rangle=
\int  \frac{{\rm d} \x}{(2 \pi)^2}  e^{-\im \q\cdot (\x-\xp)} \langle n (\q) \rangle_{th}
\label{gamma}
\end{equation}
where $\langle n (\q) \rangle_{th} $ denotes the expectation value of the photon number in mode $\q$ 
in the thermal state,
and 
we implicitly used the hypothesis of
translational invariance of the source. In particular,
the following factorisation property holds \cite{bib10}:
\begin{eqnarray}
 \langle : a^\dagger (\x) a (\xp) a^\dagger (\xpp) a (\xppp) : \rangle =
\langle  a^\dagger (\x) a (\xp) \rangle \langle  a^\dagger (\xpp) a (\xppp) \rangle 
  +
\langle \, a^\dagger (\x) a (\xppp) \rangle \langle a^\dagger (\xpp) a (\xp) \,  \rangle \; ,
\label{gamma2}
\end{eqnarray}
where $ : \, : $ indicates normal ordering.
A field with these properties is described by a thermal-like density
matrix of the form
\begin{equation}
\rho_{th} = \prod_{\q}\left\{
\sum_{m=0}^{\infty} \frac {[\langle n (\q) \rangle_{th} ]^m} {[1 + \langle n (\q) \rangle_{th} ]^{m+1} } 
\left| m,\q \right\rangle
\left\langle m,\q \right|  \right\} \: ,
\end{equation}
where $\left| m,\q \right\rangle$ denotes the Fock state with $m$ photons in mode $\q$.\\
In both the PDC and thermal case, the two outgoing beams 
travel through two distinct imaging systems,
described by their impulse response functions $h_1 (\x,\xp)$, $h_2 (\x,\xp)$ (see Fig. \ref{fig2}).
Arm 1 includes an object. Beam 1 is detected either by a point-like detector $D_1$, or by
a ``bucket" detector, 
which collects all the light in the detection plane, 
in any case giving no information on
the object spatial distribution. On the other side, detector $D_2$  spatially resolves
the light fluctuations, as for example an array of pixel detectors.
The fields at the detection planes are given by
\begin{equation}
c_i (\x_i) = \int {\rm d} \xp_i h_i( \x_i,\x_i\,') b_i (\x_i\,') + L_i (\x_i)  \quad i=1,2 \, ,
                \label{cfields}
\end{equation}
where $L_1, L_2$ account for possible losses in the imaging systems,
and depend on vacuum field operators uncorrelated from $b_1, b_2$. 
Since they do not
contribute to the normally ordered expectation values that we will calculate in the following, their
explicit expression is irrelevant.
Information about the object is extracted by
measuring the spatial correlation function of the intensities detected by $D_1$ and $D_2$,
as a function of the position $\x_2$ of the pixel of $D_2$:
\begin{equation}
\langle I_1 (\x_1) I_2 (\x_2) \rangle =
\langle  c_1^\dagger (\x_1) c_1 (\x_1) c_2^\dagger (\x_2) c_2 (\x_2) \rangle \; .
\label{eq5}
\end{equation}
All the object information
is concentrated in the correlation function of intensity fluctuations:
\begin{equation}
G(\x_1, \x_2) = \langle I_1 (\x_1) I_2 (\x_2) \rangle - \langle I_1 (\x_1)\rangle \langle I_2 (\x_2) \rangle \; ,
\label{eq6}
\end{equation}
where $\langle I_i (\x_i)\rangle = \langle c_i^\dagger(\x_i) c_i(\x_i)     \rangle $ is the 
mean intensity of the i-th beam.
Since $c_1$ and $c_2^\dagger$ commute, all the terms in Eqs. (\ref{eq5}),(\ref{eq6}) are normally ordered and 
$L_1, L_2$ 
can be neglected, thus obtaining
\begin{eqnarray}
 & G(\x_1, \x_2) = \int {\rm d} \xp_1
\int {\rm d} \xpp_1
\int {\rm d} \xp_2 \int {\rm d} \xpp_2 
  h_1^*(\x_1, \xpp_1) h_1 (\x_1, \xp_1) h_2^* (\x_2, \xpp_2) h_2 (\x_2, \xp_2) \nonumber \\
&  \left[
\langle  b_1^\dagger (\xpp_1) b_1 (\xp_1) b_2^\dagger (\xpp_2) b_2 (\xp_2) \rangle -
\langle  b_1^\dagger (\xpp_1) b_1 (\xp_1)\rangle \, \langle b_2^\dagger (\xpp_2) b_2 (\xp_2) \rangle \right] \: .
\label{eq7}
\end{eqnarray}
In the thermal case, by taking into account the transformation (\ref{BS})
and that $v$ is in the vacuum state, $b_1$  and $b_2$  in Eq.(\ref{eq7}) can be simply replaced
by $r a$ and $ta$. Next, by using Eq.(\ref{gamma2}), we arrive at the final result
\begin{equation}
G(\x_1, \x_2) = |tr|^2\left| \int {\rm d} \xp_1
\int {\rm d} \xp_2  h_1^* (\x_1, \xp_1) h_2 (\x_2, \xp_2) \langle a^\dagger (\xp_1) a(\xp_2) \rangle
\right|^2 \: ,
\label{eq12}
\end{equation}
On the other hand, also in the PDC case the four-point correlation function in Eq.(\ref{eq7}) 
has special factorization properties.
As it can be obtained from Eq.(\ref{eq1}) \cite{bib8},
\begin{eqnarray}
\langle  b_1^\dagger (\xpp_1) b_1 (\xp_1)  b_2^\dagger (\xpp_2) b_2 (\xp_2) \rangle
&=&
\langle  b_1^\dagger (\xpp_1) b_1 (\xp_1)\rangle \, \langle b_2^\dagger (\xpp_2) b_2 (\xp_2) \rangle \nonumber \\
&+&
\langle  b_1^\dagger (\xpp_1) b_2^\dagger (\xpp_2)\rangle \, \langle b_1 (\xp_1) b_2(\xp_2) \rangle
\: .
\label{eq8}
\end{eqnarray}
By inserting this result in Eq. (\ref{eq7}), one obtains
\begin{equation}
G(\x_1, \x_2)  = \left| \int {\rm d} \xp_1
\int {\rm d} \xp_2  h_1 (\x_1, \xp_1) h_2 (\x_2, \xp_2) \langle b_1 (\xp_1) b_2(\xp_2) \rangle
\right|^2 \: ,
\label{eq9}
\end{equation}
where, by using relations (\ref{eq1}),
\begin{equation}
\langle b_1 (\xp_1) b_2(\xp_2) \rangle
=
\int \frac{ {\rm d} \q }{(2\pi)^2} e^{\im \q \cdot (\xp_1 -\xp_2)} U_1 (\q) V_2 (-\q)
\: .
\label{gammaentangled}
\end{equation}
There is a clear analogy between the results in the two cases. 
Apart from the numerical factor $|tr|^2$ and  the presence of $h_1^*$ instead of $h_1$,
the second order correlation $\langle a^\dagger (\x) a(\xp) \rangle $  and the function 
$\langle n(\q) \rangle_{th}$
play in Eq.(\ref{eq12}) 
the same role as the correlation function $\langle b_1 (\x) b_2(\xp) \rangle $ 
and $U_1(\q)V_2(-q)$ in Eq.(\ref{eq9}). 
Most importantly, in both Eqs.(\ref{eq12})and (\ref{eq9}) the modulus is outside
the integral, a feature that ensures the possibility of coherent imaging
via correlation function (see e.g. \cite{bib5}).
The correlation function $\langle a^\dagger (\x) a(\xp) \rangle $ governs the properties of spatial
coherence of the thermal source\cite{bib10}.
The correlation length, or  transverse coherence length 
$l_{coh}$,
is determined by the inverse of the bandwidth $\Delta q$ of the function $\langle n(\q) \rangle_{th}$.
The same holds for the correlation $\langle  b_1 (\x) b_2(\xp) \rangle $, and the
function $U_1(\q)V_2(-\q)$ in the entangled case.

\begin{figure}[h]
{\scalebox{.4}{\includegraphics*{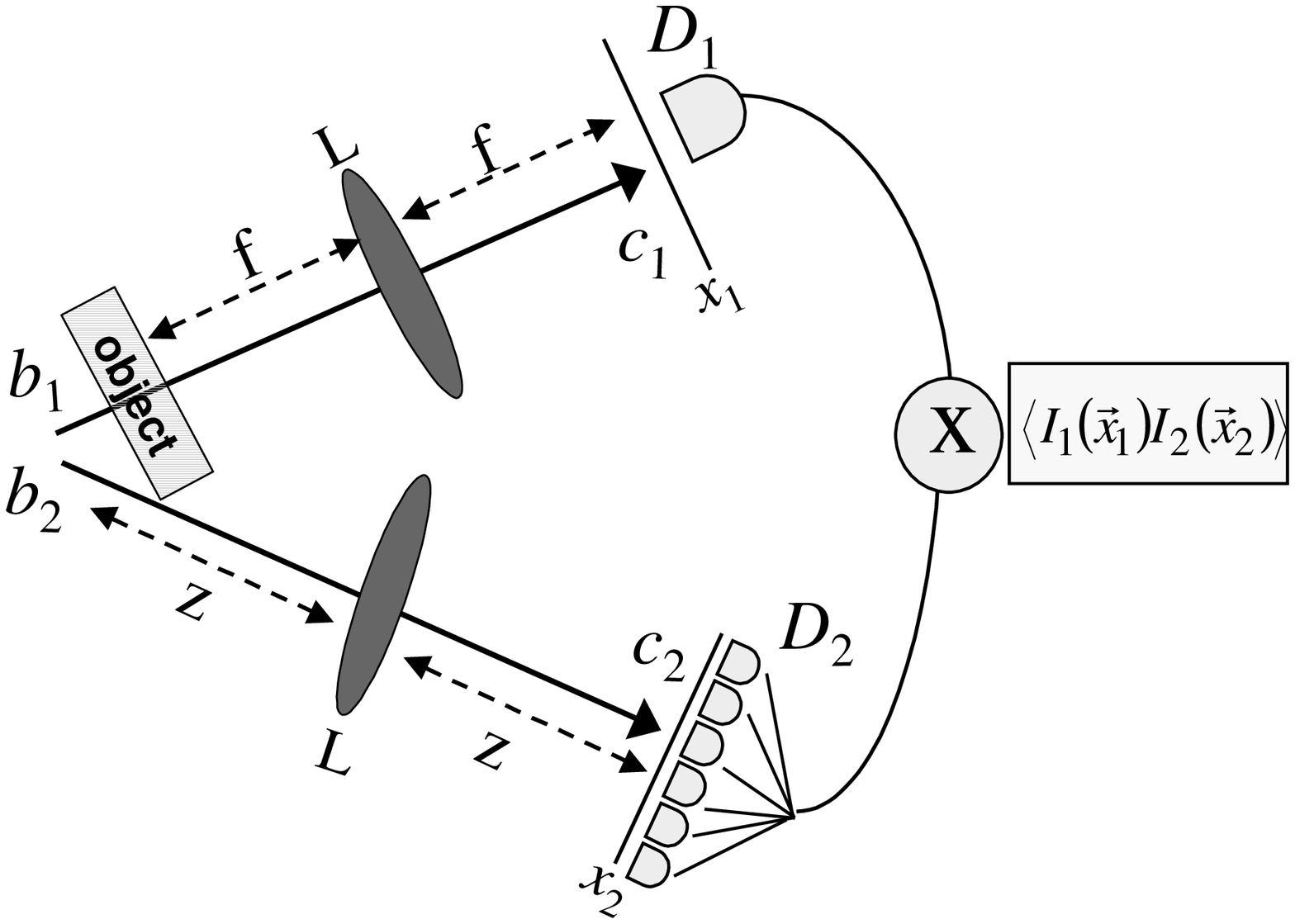}}} \caption{Imaging scheme. $L$ denotes two identical lenses of
focal length $f$. The distance $z$ is either $z=f$ or
$z=2f$.} \label{fig3}
\end{figure}

Let us now analyse two paradigmatic examples of imaging systems, mutated from the discussion of \cite{bib7}, and
described in Fig.\ref{fig3}.
In both examples the  set-up of arm 1 is fixed, and consists of an object, described by a complex
transmission function $T(\x)$, and a lens located at a focal distance $f$ from the object and from the detection
plane. Hence, 
$h_1(\x_1,\xp_1)= -\frac{\im}{\lambda f} \exp{\left(-\frac{2\pi\im}{\lambda f} \x_1 \cdot \xp_1 \right)} T(\xp_1)$,
with $\lambda$ being the wavelength. In arm 2 there is a single lens placed at a distance z both from
the source and from the detection plane 2.\\
In the first example we assume z=f (we take the two lenses identical for
simplicity), so that
$h_2(\x_2,\xp_2)= -\frac{\im}{\lambda f} \exp{\left(-  \frac{2\pi\im}{\lambda f}  \x_2 \cdot \xp_2\right) } \; .$
By inserting these propagators into Eq.(\ref{eq12}), and taking into account 
the  expression of $\langle a^\dagger (\x) a(\xp) \rangle $ given by 
Eq.(\ref{gamma}), we obtain
\begin{equation}
G(\x_1, \x_2) 
=\frac{(2\pi)^2 |rt|^2}{(\lambda f)^4}
\left| \langle n ( -\x_2 \frac{2\pi}{\lambda f} ) \rangle_{th}  \, 
\tilde T \left( (\x_2 -\x_1) \frac{2\pi}{\lambda f}  \right)\right|^2 \: ,
\label{diffpatt}
\end{equation}
where $
\tilde T (\q) = \int \frac{\rm d \x}{2\pi} e^{-\im \q \cdot \x} T(\x)$
is the amplitude of the diffraction pattern from the object. This
has to be compared with the result of the entangled case
(see Eq.7 of \cite{bib7}), where the combination $\x_2+ \x_1$
appears instead of $\x_2 -\x_1$ and $ U_1 V_2 $ instead of
$\langle n \rangle_{th}$. In both cases, the diffraction pattern
from the object can be reconstructed, provided that 
the spatial
bandwidth $\Delta q$ is larger than the maximal $q$ vector
appearing in the diffraction pattern, or, equivalently, provided that 
$l_{coh} < l_{o}$, where $l_{o}$ is the smallest scale of variation
of the object spatial distribution. Best
performances of the scheme are achieved for
spatially incoherent light, $l_{coh} \to 0$. We 
remind the well-known result that, when $l_{coh} < l_o$, no
information about the diffraction pattern of the object is
obtained by  detection of light intensity distribution in arm 1 by an array of pixels.\\
In the second example, we set z=2f, so that $h_2 (\x_2,\xp_2) = \delta (\x_2+\xp_2) \exp{\left(-\im |x_2|^2
\frac{\pi}{\lambda f}\right)}$.
Inserting this in Eq. (\ref{eq12}),  we get:
\begin{eqnarray}
G (\x_1, \x_2) &=&
\frac{|rt|^2}{ (\lambda f)^2}
\left|
\int \frac{\rm d \q}{2\pi}
 \left\langle n(\q) \right\rangle_{th}
\tilde T \left( \x_1  \frac{2\pi}{\lambda f } -\q  \right) e^{\im \q \cdot \x_2}
\right|^2
\label{image1} \\
&\approx& 
\frac{|rt|^2}{(\lambda f)^2} 
\left| \langle n( \x_1\frac{2\pi}{\lambda f} )   \rangle_{th}\right|^2
  \left|  T \left( -\x_2  \right)
\right|^2 \: ,
\label{image2}
\end{eqnarray}
where in the second line $l_{coh} < l_{o}$ was assumed, so that 
$\langle n(\q)\rangle_{th}$ is roughly constant in the region of $\q$ plane where the
diffraction pattern does not vanish, and it can be taken out from the integral in (\ref{image1}).
In this example the correlation function provides information about the image of the
object.  A similar result holds for the case of entangled beams (see Eq.(8) of \cite{bib7}).

Our results appear suprising, if one has in mind the case of a coherent beam impinging on a beam splitter,
where the two outgoing fields are uncorrelated, i.e. $ G(\x_1, \x_2) =0$.
However, when the input field is an intense thermal beam, i.e. the  photon number per mode 
is not too small,
the two outgoing field are well correlated in space. To prove this point,
let us consider the number of photons detected in
two small identical portions  R (``pixels") of the beams in the near field    
immediately after the beam splitter,  $N_i= \int_{R} {\rm d} \x \, b_i^\dagger(\x) b_i (\x)\:$ $i=1,2$, 
and the difference
$N_-= N_1-N_2$. Making use of the transformation (\ref{BS}),
it can be proven that,  for  $|r|^2= |t|^2=1/2$, 
the variance 
$\langle \delta N_-^2 \rangle = \langle N_-^2\rangle -\langle N_-\rangle^2 $ is given by
\begin{equation}
\langle \delta N_-^2\rangle=
\langle N_1\rangle+ \langle  N_2\rangle\: ,
\label{SN}
\end{equation}
which corresponds exactly to the shot noise level. On the other side, by using the identity
$\langle \delta N_-^2\rangle=
\langle \delta N_1^2\rangle  +  \langle \delta N_2^2\rangle- 2\langle \delta N_1 \delta N_2 \rangle$
and taking into account that 
$\langle \delta N_1^2\rangle=\langle \delta N_2^2\rangle$ for $ |r|^2=|t|^2$, the normalized correlation is  
given by
\begin{equation}
C\stackrel{{\rm def}}{=}\frac{\langle \delta N_1 \delta N_2 \rangle }{\sqrt{\langle \delta N_1^2\rangle} 
\sqrt{ \langle \delta N_2^2\rangle} }
=1-\frac{\langle  N_1\rangle}{  \langle \delta N_1^2\rangle   } \; .
\label{C}
\end{equation}
For any state $0\le|C| \le 1$, where the lower bound corresponds to the coherent state level,
and the upper bound is imposed by Cauchy-Schwarz inequality. For the thermal state,
provided that the pixel size is not too large with respect to $l_{coh}$,
$\langle \delta N_1^2\rangle  \approx \langle N_1 \rangle  + \langle N_1 \rangle ^2  $,
so that the correlation (\ref{C}) never vanishes.
For thermal systems with a large number of photons  
$ \langle N_1\rangle   / \langle \delta N_1^2\rangle  \approx
1/ \langle N_1 \rangle   <<1$, and $C$ can be made close its maximum value.
Even more important,
it is not difficult to show that  Eqs.(\ref{SN}) and (\ref{C}) hold
in any plane linked to the near field plane by a Fresnel transformation, in the absence of losses. 
In particular, a high level of
pixel-by-pixel correlation can be observed in the far-field plane. 
We remark that, despite 
$C$ can be made close to 1 by increasing the mean number of photons, the correlation never
reaches the quantum level, as shown by Eq.(\ref{SN}). 

For the entangled beams
produced by PDC, spatial correlation is present both in the near and in the far field, with
the ideal result $\langle \delta N_-^2\rangle= 0$, $C=1$ in both planes \cite{bib8}. In this case,
the far field correlation is between symmetric pixels, due to the different way in which the four-point
correlation function factorises.
In \cite{bib7} we analysed the effect of replacing the pure PDC entangled state with two mixtures that 
exactly preserve
the spatial signal-idler quantum correlations, either in the far or in the near field. It turned out
that, when considering the ``far-field mixture",
 the pure state results in the z=f configuration of Fig.\ref{fig3} can be reproduced, but there 
is no information about the image in the z=2f configuration. The converse is true considering the ``near-field mixture". 
The two beams generated by splitting thermal light 
are instead imperfectly correlated both in the near and in the far-field. However, by using intense thermal light, classical intensity correlation is strong enough to reproduce
qualitatively the results of both the z=f and the z=2f configuration. \\ \indent
A complete comparison of the performances in the classical and quantum  regimes
requires extended numerical investigations, 
describing realistic thermal sources,
which is outside the scope of this paper.
A key role in this comparison will be  played by the issue of the visibility of the information, which
is retrieved by subtracting from the measured correlation function (\ref{eq5}), the background  term 
$ \langle  I_1 (\x_1) \rangle \langle I_2 (\x_2)\rangle $ (see Eq.(\ref{eq6})).
A measure of the visibility is given by ${\cal V}= \frac{G(\x_1,\x_2)}{\langle  I_1 (\x_1) I_2 (\x_2)\rangle }$.\\
A first 
remark concerns the presence of $\langle n(\q)\rangle_{th}$ in Eq. (\ref{eq12}) in place of 
$ U_1(\q) V_2(-q)$ in Eq.(\ref{eq9}). As a consequence,
in the thermal case $ G(\x_1, \x_2)$ scales
as $\langle n(\q)\rangle_{th}^2$, while 
in the entangled case, it scales as 
$\left|U_1(\q) V_2(-q)\right|^2= \langle n(\q)\rangle + \langle n(\q)\rangle^2 $,
where $\langle n(\q)\rangle =|V_2(-\q)|^2= |V_1(\q)|^2$ is 
the mean number of photons per mode in the PDC beams,
and
$|U_1(\q)|^2 = 1+|V_1(\q)|^2$ (see e.g \cite{bib8}). 
This difference is immaterial when $\langle n(\q)\rangle >> 1$,
while it becomes relevant for a small photon number, because the background 
$\langle  I_1 (\x_1) \rangle \langle I_2 (\x_2)\rangle 
\propto \langle n(\q)\rangle^2$ is negligible with respect to $G(\x_1,\x_2) \propto \langle n(\q)\rangle $. 
Hence, in the regime of single photon pair detection,
the entangled case presents a much better visibility of the information
with respect to  classically correlated thermal beams (see also \cite{Klyshko}).\\
A second remark concerns the role of the temporal argument. Calculations 
that will be reported elsewhere 
show that the visibility scales as the ratio between the coherence time of the source $\tau_{coh}$
and the detection time  (see also \cite{bib8b,bib11}). 
In practical applications, 
this implies that conventional thermal sources, with very small coherence times, are not suitable
for the schemes studied here. A suitable source should present a relatively long coherence time,
as for example a sodium lamp, or the chaotic light produced by scattering a laser beam through
a random medium \cite{bib15}. As a special example of a thermal source, one can
 consider the signal field or the idler field generated by PDC.
\footnote{This opens new possibilities, offered by the combination of the correlated imaging from entangled beams
and from classically correlated beams.}
\begin{figure}[ht]
{\scalebox{.6}{\includegraphics*{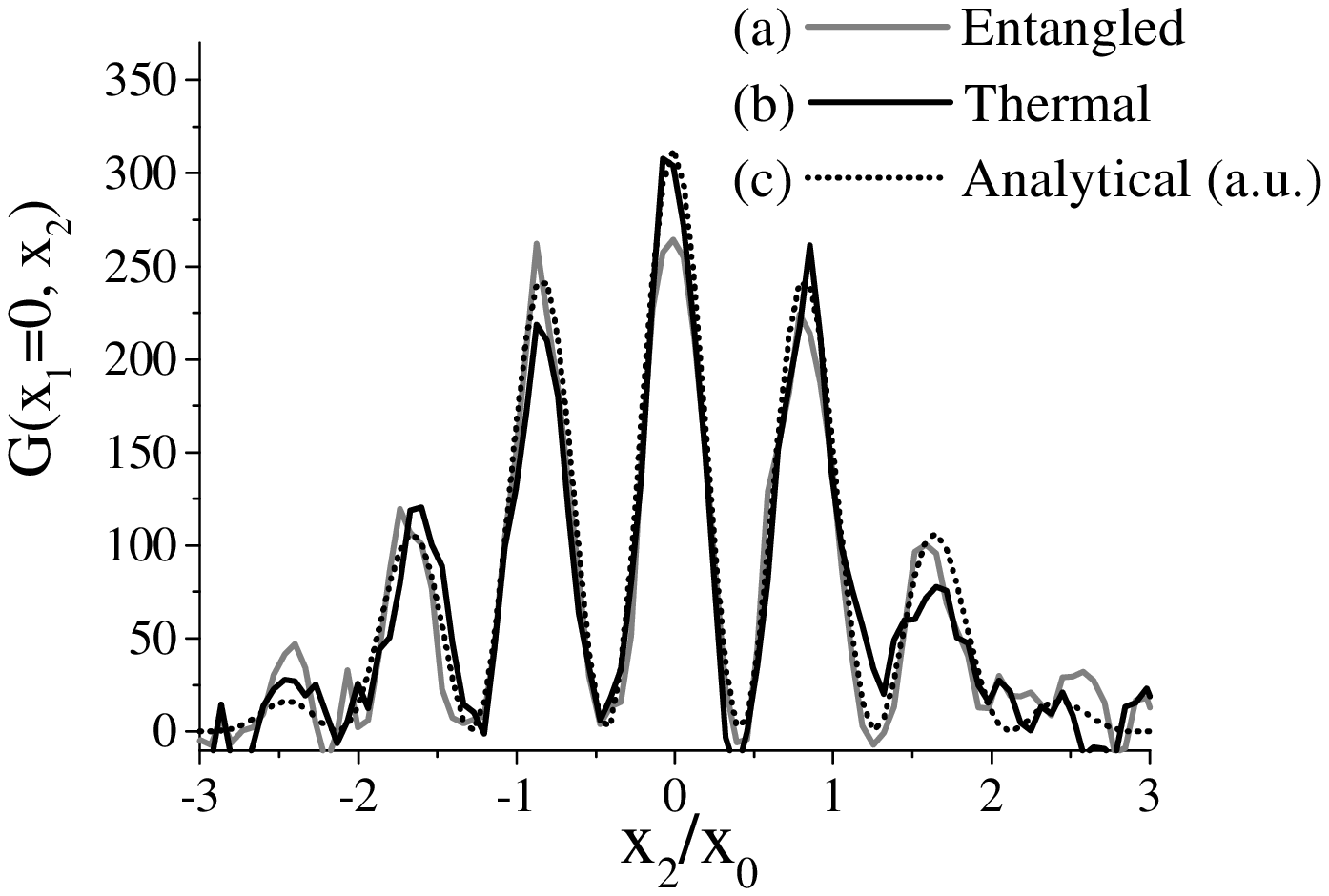}}} \caption{Numerical simulation of the reconstruction of 
the  diffraction pattern of a double slit in the scheme z=f of Fig.\ref{fig3}.
$G(\x_1,\x_2)$ versus $\x_2$,  after $10000$ shots for
(a) entangled signal/idler beams from PDC, (b)
classically correlated beams by splitting the idler beam. (c) is the analytical 
result of Eq.(\ref{diffpatt}). Parameters
are those of a 4 mm $\beta$-barium-borate crystal ($l_{coh}=  16.6 \mu$m, $\tau_{coh}=0.97$ps).
The pump waist is $664 \mu$m, and the pulse duration is $1.5 ps$. $x_0= \Delta q \lambda f/(2\pi)$} \label{fig4}
\end{figure}
Fig. \ref{fig4} shows the results of a numerical simulation
for the reconstruction of the diffraction pattern of a double slit,
in the scheme z=f of Fig.\ref{fig3}. It compares
the use of signal/idler 
entangled beams (curve (a)), and two classically correlated beams obtained by symmetrically
splitting the idler beam (curve (b)). Parametric gain is such  that
$\langle n (\q) \rangle \approx 750$ at its maximum for (a),
and $\langle n (\q) \rangle \approx 1500$ for (b), so that the the mean photon number of beams $b_1$ $b_2$
is the same in the two simulations.
This example not only shows that in the regime of high photon number the quantum and classical correlation offer
similar performances, but also that although the visibility is rather poor, ${\cal V} \approx 0.05$ in both cases,
this is enough to correctly retrieve the desired information after a reasonable number of pump shots. The numerical
data for the image reconstruction (scheme z=2f of Fig.\ref{fig3}) confirm these results.
\\ \indent
In conclusion, we have shown a deep analogy between the use of entangled beams and classically correlated
beams from a thermal source in imaging schemes based on correlation measurements. As it was already recognized in other
contexts (see e.g. \cite{Franson}), in the small photon number regime, a definite advantage of the quantum configuration 
is represented by a better visibility.\\
Our result, that is possible to perform
coherent imaging without spatial coherence by
using thermal light in combination of a beam splitter is reminiscent of the Hanbury-Brown and Twiss
interferometric method for determining the stellar diameter\cite{bib16},
or of the detection of  the fringes arising from interference of two independent thermal sources \cite{bib17}.
However, here, we define a technique to achieve a full coherent imaging with a great deal of flexibility. 
Since the required correlation is classical,
a high quantum efficiency of detectors is not necessary.
\acknowledgments{This work was carried out in the framework of the FET project QUANTIM of the EU.
We are grateful to S. Sergienko, B. Boyd and E.Lantz for  stimulating discussions.}


\begin{references}
\bibitem{bib1} A.V.~Belinsky and D.N.~Klyshko, Sov.~Phys JETP {\bf 78}, 259 (1994).
\bibitem{bib2} D.V. Strekalov, A.V. Sergienko, D.N. Klyshko and Y.H. Shih, Phys. Rev. Lett. {\bf 74}, 3600 (1995).
\bibitem{bib3} 
T.B. Pittman, Y.H. Shih, D.V. Strekalov and A.V. Sergienko,
               Phys. Rev. A {\bf 52}, R3429 (1995).
\bibitem{bib4} P.H.S. Ribeiro, S.Padua, J.C.Machado da Silva and G.A. Barbosa,
               Phys. Rev. A {\bf 49} 4176 (1994).
\bibitem{bib5} A.F. Abouraddy, B.E.A. Saleh, A.V. Sergienko, and M.C. Teich
               Phys.~Rev.~Lett.~{\bf 87}, 123602 (2001);
              J. Opt. Soc. Am. B {\bf 19}, 1174 (2002).
\bibitem{bib7} A.Gatti, E. Brambilla,L.A. Lugiato, Phys.Rev.Lett. {\bf 90}, 133603-1 (2003)
\bibitem{bib13} R.S. Bennink, S.J. Bentley and R.W. Boyd,
            Phys. Rev. Lett {\bf 89}  113601 (2002).
\bibitem{bib14} M. D'Angelo and Y.H. Shih, e-print quanth-ph/0302146
            (2003).
\bibitem{bib8b} B.E.A~Saleh,A.F.~Abouraddy, A.V~Sergienko and M.T.~Teich, Phys. Rev. A {\bf 62}, 043816 (2000).
\bibitem{bib8} E. Brambilla, A. Gatti, M. Bache and L.A. Lugiato, submitted; e-print quanth-ph/0306116.
\bibitem{bib9} M.I.~Kolobov, Rev.~Mod.~Phys. {\bf 71}, 1539 (1999).
\bibitem{bib10} L. Mandel and E. Wolf, {\it Optical Coherence and Quantum Optics}, (Cambridge University Press, 1995).
\bibitem{bib11} A. Gatti, E. Brambilla, L. A. Lugiato, SPIE Proceedings vol.{\bf 5161} (2003), to appear.
\bibitem{bib15}W. Martiessen and E. Spiller, Am. J. Phys. {\bf 32}, 919 (1964).
\bibitem{Klyshko}A.V.~Belinsky and D.N.~Klyshko, Phys.~Lett. A {\bf 166}, 303 (1992).
\bibitem{Franson}Z.Y. Ou and L. Mandel, J. Opt. Soc. Am B {\bf 7},2127 (1990).
\bibitem{bib16} R. Hanbury-Brown and R.Q. Twiss, Nature (London) {\bf 177}, 27 (1956).
\bibitem{bib17} S.J.~Ku, D.T.~Smithey and M.G.~Raymer, Phys.Rev.A {\bf 43}, 4083 (1991).
\end{references}
\end{document}